\documentclass[10pt,twocolumn,twoside]{article}

\usepackage[utf8]{inputenc}
\usepackage[T1]{fontenc}
\usepackage[english]{babel} 
\usepackage{CJKutf8} 

\usepackage{graphicx}
\usepackage{amsmath,amssymb,amsfonts}
\usepackage{float}
\usepackage{xcolor}
\usepackage{titlesec}
\usepackage{authblk} 
\usepackage{fancyhdr}
\usepackage{abstract}
\usepackage{tipa} 
\usepackage[numbers,sort&compress]{natbib}
\usepackage{orcidlink}
\usepackage{hyperref}

\definecolor{pnasblue}{RGB}{0, 114, 188}
\hypersetup{colorlinks=true, linkcolor=pnasblue, citecolor=pnasblue, urlcolor=pnasblue}

\usepackage[left=1.5cm,right=1.5cm,top=2cm,bottom=2cm]{geometry}

\titleformat{\section}{\large\bfseries\color{pnasblue}\sffamily}{\thesection}{1em}{}
\titleformat{\subsection}{\normalsize\bfseries\sffamily}{\thesubsection}{1em}{}

\title{\huge \bfseries \sffamily  The Collapse of Multilayer Predation and the Emergence of a Monolithic Leviathan  \\ \Large \color{gray} \huge \textnormal{多层掠夺的崩塌和单一利维坦的崛起}}

\author[a,b,1]{Li Tuobang 李拓邦 \orcidlink{0000-0002-2257-2603}}
\affil[a]{Independent Researcher, Zhaoqing, China 独立研究员，肇庆，中国}
\affil[b]{University of California, Berkeley, USA 加州大学伯克利分校，美国}

\affil[1]{To whom correspondence should be addressed. 联系方式。E-mail: lituobang@hotmail.com}

\newcommand{\acknow}[1]{\section*{Acknowledgment 致谢} {\small #1}}
\newcommand{\dataavail}[1]{\section*{Data Availability 数据可获取} {\small #1}}

\begin{document}

\begin{CJK*}{UTF8}{gbsn}

\twocolumn[
  \begin{@twocolumnfalse}
    \maketitle
    \begin{abstract}
    This paper constructs a multilayer recursive game model to demonstrate that in a rule vacuum environment, hierarchical predatory structures inevitably collapse into a monolithic political strongman system due to the conflict between exponentially growing rent dissipation and the rigidity of bottom-level survival constraints. We propose that the rise of a monolithic political strongman is essentially an "algorithmic entropy reduction" achieved through forceful means by the system to counteract the "informational entropy increase" generated by multilayer agency. However, the order gained at the expense of social complexity results in the stagnation of social evolutionary functions.

    本文构建了一个多层递归博弈模型，论证了在规则真空环境下，多层级掠夺结构由于租金耗散的指数级增长与底层生存约束的刚性冲突，必然向单一政治强人体制收缩。我们提出，单一政治强人的兴起本质上是系统为了对抗多层级代理所产生的“信息熵增”，通过强力手段实现的算法化减熵。然而，这种通过牺牲社会复杂性换取的秩序，其代价是社会演化功能的停滞。
    
    \end{abstract}
    
    \vspace{0.5cm}
    {\small \color{gray} 本文英文版已在《国际社会科学研究与评论杂志》，2026年第1期上发表。The English Version of this paper has been publish in International Journal of Social Science Research and Review.
\href{https://ijssrr.com/journal/article/view/3198}{https://ijssrr.com/journal/article/view/3198}}
    \vspace{1cm}
  \end{@twocolumnfalse}
]

\section{Introduction 引言}

在经典制度经济学与政治哲学的叙事框架中，利维坦（Leviathan）的兴起通常被解读为社会主体为了摆脱“每个人对每个人的战争”这一霍布斯陷阱（Hobbesian Trap）而达成的理性契约决策 \cite{hobbes1651leviathan}。然而，这种传统的“社会契约论”往往预设了一个简化模型，即掠夺行为发生在同一物理平面上的原子化个体之间。

本文提出，当社会处于真实的“规则真空”状态时，最初的掠夺博弈并非平面化的简单冲突，而是具有显著的递归性与嵌套性。在这种结构下，社会迅速分化为层级化的权力链条：一个掠夺者在向下实施价值榨取的同时，往往也成为更高层级掠夺者的目标。这种“食肉动物亦是猎物”的递归链条，构成了本文所定义的“多层级掠夺结构”。

本文的核心论点在于：多层级的掠夺结构在数学与信息学意义上是不稳定的。 首先，根据租金耗散理论，多层嵌套的掠夺会导致社会总剩余在复杂的层级博弈中被迅速抵消 \cite{tullock1967welfare}。 其次，从信息论的视角来看，每一层掠夺层级都充当了一个具有噪声的信息通道。由于中间代理人为了最大化自身租金，具有天然的动机去操纵向上汇报的产出数据和向下传递的压力信号，这导致系统内的信息熵随层级增加呈指数级爆表 \cite{shannon1948mathematical}。

当信息熵达到临界点时，最高层级的掠夺者（委托人）将彻底丧失对底层生产者（代理人）生存状态的观测能力，导致底层参与者极易跌破其“生存约束”的红线，从而引发整个社会生产系统的连锁坍塌。

因此，单一政治强人的兴起，本质上是复杂系统为了对抗“多层级递归掠夺”所带来的高熵乱局，而内生地产生的一种“算法化减熵”过程。强人通过暴力手段强行平抑中间层级的自主性，将高维度的混乱博弈压缩为单向的低熵指令，以实现对社会资源更加“精准且可持续”的榨取。

\section*{霍布斯陷阱：规则真空下的原子化掠夺}

在探讨复杂的递归层级之前，我们假设在一个法律体系约束为零的初始环境中，社会由 $N$ 个原子化的经济主体组成。

\subsection*{收益矩阵与理性选择的悖论}

每个参与者 $j \in N$ 面临生产（$P$）与掠夺（$G$）的抉择。设 $\alpha$ 为生产的基础回报，$\beta$ 为掠夺动作能攫取的盈余。

在规则真空下，博弈的收益矩阵如下：$$\begin{array}{c|cc}
& P (\text{生产者}) & G (\text{掠夺者}) \\
\hline
P (\text{生产者}) & (\alpha, \alpha) & (0, \alpha + \beta) \\
G (\text{掠夺者}) & (\alpha + \beta, 0) & (-C, -C)
\end{array}$$\begin{itemize}\item 

(P, P) 的脆弱性： 虽然双方生产能实现社会福利最大化 $2\alpha$，但在 $L=0$ 时，参与者无法做出可信承诺。\item 

(G, G) 的必然性： 对于任何一方，如果对方选择 $P$，我选 $G$ 收益更高（$\alpha + \beta > \alpha$）；如果对方选择 $G$，我若选 $P$ 收益为 $0$，而选 $G$ 虽然损失 $C$，但由于 $-C$（冲突损耗）通常被视为避免被彻底奴役或杀死的代价，个体在博弈中会陷入这种“安全陷阱”。\end{itemize}

由于 $\alpha + \beta > \alpha$，掠夺（G）是个体的严格占优策略。系统无可避免地向 $(G, G)$ 收敛。在这个均衡点，社会总产出从 $2\alpha$ 跌落至 $-2C$。这解释了为什么在规则真空下，社会无法自发产生繁荣。这种状态就是典型的霍布斯陷阱（Hobbesian Trap） \cite{hobbes1651leviathan}。

\subsection*{租金耗散与信息熵的极大化}

在 $(G, G)$ 的纳什均衡下，社会陷入了全额的“租金耗散” \cite{tullock1967welfare}。$-C$ 代表了社会资源被无效空耗。这包括：

\begin{itemize}\item 
防御性支出： 建造围墙、购置武器。\item 
生命与资本毁损： 暴力冲突导致的生产资料破坏。\item 
机会成本： 原本可用于研发和生产的劳动力被吸纳进非生产性的掠夺活动。\end{itemize}

\subsubsection*{信息论视角：信号消失与熵增}

从信息论视角看 \cite{shannon1948mathematical}，这一状态代表了系统信息熵的极大化：\begin{itemize}\item

高噪声环境： 当每个人都在掠夺和欺诈时，系统中充斥着虚假信号（伪装生产能力或伪装暴力能力）。\item 

互信息的归零： 参与者之间的互信息（Mutual Information）极低，无法达成任何协同博弈。\item 

系统热寂： 系统内能被转化为无序的内部摩擦。由于没有稳定的分配规则，社会无法积蓄负熵来形成更高级的组织形式。\end{itemize}

\subsection*{演化动力：为什么 $(G, G)$ 会导致层级化？}

大规模的$(G, G)$ 冲突是不可持续的，因为它会导致弱者的物理消亡。为了跳出收益为负的 $-C$ 状态，博弈开始演化：

\begin{enumerate}\item 

规模效应突破： 只要有若干人组成“暴力联盟”，他们就能以更低的平均 $C$ 压制原子化的 $P$，从而获得 $\alpha + \beta$。

\item 

自愿奴役： 每个人的$-C$不一样，产出$\alpha$也不一样。对有些人来说，$-C$是不可接受的。弱者为了避免 $-C$ 带来的彻底毁灭，愿意接受被剥夺大部分产出的条件以换取生存。\end{enumerate}

这种从“原子化冲突”向“层级化掠夺”的跨越，正是本文接下来要讨论的多层递归层级的起点，也就是利维坦的兴起。

\section*{多层掠夺模型的设定}

为了量化规则真空下的系统演化，我们构建一个基于递归博弈的多层级信息-价值交换模型。

\subsection*{递归榨取结构}

假设社会系统由 $k$ 个相互嵌套的权力层级（Power Hierarchy）组成。在这种结构中，价值流从底层向上汇聚，而指令流从顶层向下传导：

\begin{itemize}\item 

底层生产者 ($L_0$)：系统能量的唯一供给方，其初始产出为 $\alpha$。在规则真空下，$L_0$ 缺乏组织化议价能力，其产出被视为可无偿提取的“公共池资源”（Common-pool Resources） \cite{ostrom1990governing}。

\item 

中间层 ($L_1 \dots L_{k-1}$)：具有典型的“双重代理”属性（Double Agency）。根据对层级官僚博弈的研究，中间层级并非简单的传声筒，而是具有独立目标函数的“驻扎型掠夺者” \cite{olson1993dictatorship}。他们既面临来自上层 $L_{i+1}$ 的榨取压力，又拥有对下层 $L_{i-1}$ 的自由裁量榨取权。

\item 

顶层强人 ($L_k$)：系统的剩余索取者。在 $k > 1$ 的混沌状态下，$L_k$ 仅仅是名义上的统治者，其真实的租金提取率受到中间层级“信息割据”的严重限制。\end{itemize}

\subsection*{信息熵与生存摩擦}

在每一对相邻层级 $(L_i, L_{i-1})$ 之间，存在着不可消除的信息不对称。由于缺乏市场价格信号和法治化的审计机制，掠夺者必须支付监控成本 $C_m$ 以降低代理风险 \cite{jensen1976theory}。

\subsubsection*{1. 代理熵（Agency Entropy）的累积}

我们定义中间层级的行为为“代理熵增”。在规则真空下，监督契约失效，中间层级 $L_i$ 为了最大化自身剩余，采取对称性的瞒报策略：

\begin{enumerate}\item 

向上： 通过增加观测噪声，隐匿底层的真实生产潜力，以降低上缴配额。

\item 

向下： 补偿性压榨。由于 $L_i$ 自身的租金被 $L_{i+1}$ 提取，为了维持其生存约束，它会过度利用其对底层的暴力垄断权，将压榨强度推向极限。\end{enumerate}

从信息论视角看，这种双向寻租（Double-sided Rent-seeking）导致了信道容量（Channel Capacity）的剧烈收缩 \cite{cover1999elements}。当层级 $k$ 增加时，系统信噪比（SNR）呈指数级下降。

\subsubsection*{2. 生存摩擦与硬约束}

每一层级的榨取行为都受到底层物理极限的限制。在多层递归下，由于中间层级互不通气，他们对底层的探测表现为一种“无协调的盲目刺探”。每一次榨取都是在触碰底层的物理红线。当代理熵导致的信号扭曲使得顶层 $L_k$ 误认为底层仍有结余时，最终下达的掠夺指令将成为穿透生存底线的最后一根稻草。

\section*{多层掠夺的坍塌逻辑}

多层级掠夺体系的结构性矛盾在于：掠夺者对租金的贪婪是线性级的，但系统因层级嵌套产生的内耗却是指数级的。这种不对称性决定了该结构存在一个物理学上的生存边界。

\subsection*{租金耗散与“递归产出”的级联效应}

在递归博弈中，底层的生存空间受到各层级掠夺率 $r_i$ 的连锁挤压。底层生产者 $L_0$ 最终能保留的有效资源 $Y_{net}$ 遵循级联衰减律：$$Y_{net} = \alpha \cdot \prod_{i=1}^{k} (1 - r_i)$$

这种乘法效应意味着，即使每一层级的掠夺率 $r_i$ 处于适度区间，随着层级 $k$ 的增加，$Y_{net}$ 也会迅速趋于零。

\subsubsection*{1. 生存约束的脆弱性与噪声放大}

在规则真空环境下，各层级掠夺者之间处于“非合作博弈”状态，缺乏统一的税收协调机制 \cite{ostrom1990governing}。\begin{itemize}\item 

随机干扰因子：每一层的 $r_i$ 实际上包含了一个随机波动项 $\epsilon_i$，代表了中间层级的临时性贪婪或对 $I_{ext}$ 的误读。

\item 

穿透效应：由于缺乏法治和对 $r_i$ 总和的硬性约束，层级 $k$ 越多，总提取率偏离底层承载能力的方差就越大。这种“无协调掠夺”会导致底层留存收益频繁且不可预测地穿透物理生存线 $\Phi_j(t)$。\end{itemize}

\subsection*{熵爆表导致的系统自毁}

根据 \cite{tullock1967welfare} 的理论，租金耗散（Rent Dissipation）不仅包含了直接的价值转移，更包含了为了维持这种转移而付出的社会资源。在多层结构中，这种耗散被进一步放大。

\subsubsection*{1. 代理成本的凸性增长}

层级间的代理成本 $C(r_i, \Delta H)$ 取决于该层的信息熵 $\Delta H$。每一层为了防止下层瞒报，必须建立独立的暴力监控机构。根据控制论中的“管理幅度”原理，监控成本随层级 $k$ 的增加呈典型的凸函数增长 \cite{williamson1967hierarchical}：$$C_{total}(k) = \sum_{i=1}^{k} \exp(\theta \cdot i)$$

其中 $\theta$ 为信息损耗系数。

\subsubsection*{2. 收益归零律}

当系统层级扩张到临界点 $k^*$ 时，为了维持金字塔稳定所需的监控总成本将超过系统能榨取的总租金。此时，系统的边际收益为负：$$\lim_{k \to \infty} R_{total} = \left( \alpha - Y_{net} \right) - \sum_{i=1}^{k} C(r_i, \Delta H) \le 0$$

这种状态在信息论中被称为“信息热寂”，在制度经济学中则表现为系统的物理性坍塌。此时，中间层级会因为无法获得足够的“分红”而倒戈，或者底层生产者由于生理再生产中断而集体消亡，迫使系统必须经历一次剧烈的“减熵重组”。

而且这个过程随着掠夺方式的不断传播，技术生产力的推动，正在变得越来越快。在人类社会早期，这个过程可能长达数百年，而到最近两千年，这个过程通常只有不到一百年的时间。在工业革命后更是缩短到三十年左右\cite{tainter1988collapse}。

\section*{单一强人的内生与减熵过程}

多层递归掠夺引发的熵增终将触及系统崩溃的临界点。为了防止系统进入物理意义上的“热寂”状态，博弈演化逻辑会驱动权力结构发生剧烈的相位转变，即向单一强人（Monolithic Strongman）体制收缩。

\subsection*{垂直整合的效率必然与信息压缩}

单一强人的兴起并非简单的暴力夺权，而是系统为了消除“多层级代理成本”而进行的垂直整合。根据产业组织理论，垂直整合可以有效消除“双重边际化”（Double Marginalization）导致的福利损失 \cite{williamson1985economic}。

\subsubsection*{1. 中间层级的“功能阉割”与去中介化}

在 $k > 1$ 的多层结构中，每一个中间层级都是一个独立的“租金中心”。单一强人通过暴力手段实现的减熵，核心在于对中间层级的功能阉割：

\begin{itemize}\item 

权力剥夺： 将原本拥有独立掠夺权的“小强人”降维为纯粹的、仅具备执行职能的代理人。\item

信息压缩： 强人通过建立垂直的指令通道，绕过中间层的层级阻隔。这种“去中介化”大幅降低了系统的信息传输熵。\end{itemize}

\subsubsection*{2. 针对 $I_{ext}$ 的精准榨取模型}

当层级收缩至 $k=1$ 时，强人垄断了对外生信息 $I_{ext}$ 的刺探权与解释权。这种垄断使得掠夺不再是随机的打击，而成为了算法化的精准榨取。强人的最优策略变为解如下带约束的优化问题：$$\max_{r} \Pi = r \cdot \alpha(r) - C(r, H)$$$$\text{s.t. } (1-r)\alpha \ge \Phi_j(t) + \epsilon$$

其中，$H$ 代表压缩后的系统熵。与多层掠夺不同，单一强人具有保护生产底线 $\Phi_j(t)$ 的内在激励，因为他是“驻扎型掠夺者”，必须保护其长期租金流的承载基座 \cite{olson1993dictatorship}。

\subsection*{强人作为“谢林点”下的绝望共识}

在规则真空产生的 $-C$ 冲突（即多头掠夺引发的军阀混战）中，社会各阶层的预期是极度紊乱的。此时，单一强人的存在成为了博弈论意义上的谢林点（Schelling Point） \cite{schelling1980strategy}。

\subsubsection*{1. 低熵指令的协调功能}

相比于多层掠夺下各派系互不兼容、频繁变动的掠夺标准，单一强人的指令虽然严酷，却具有确定性。在博弈者看来，接受一个已知的、唯一的掠夺标准，其期望损耗远低于应对多个未知的、不可预测的掠夺者的总损耗。

\subsubsection*{2. 绝望的共识}

这种向强人聚拢的行为并非出于对权力的崇拜，而是出于对高熵混乱的极度恐惧。社会各阶层（包括部分中间掠夺层）为了寻求保护和稳定的生产预期，会达成一种“绝望的共识”，主动将其自主权让渡给单一强人。这种减熵过程虽然剥夺了社会的演化空间，却在短期内修复了生存线穿透的风险。

\section*{定期清洗和打击绝对贫困：利维坦的动态去熵与原子化校准}

尽管强人统治在结构上表现为单一中心，但在实际运行中，利维坦仍需依赖一个中间阶层（官僚体系、特务机构或代理人集群）来行使意志并提取租金。然而，这种依赖性构成了强人的热力学悖论：中间阶层在执行职能的过程中，会不可避免地利用信息不对称和行政权力进行局部能量吸积，形成新的、非正式的次级权力集群。

为了防止这些次级集群演化为能够挑战中心的多层结构，强人必须引入定期清洗机制。

1. 抑制局部吸积的正反馈过程

在演化博弈论中，中间阶层的代理人倾向于通过结盟和寻租来降低自身的博弈风险\cite{olson2022rise}。如果不加干预，这些局部链接会迅速增强，产生“局部负熵”并形成半自主的权力堡垒。定期清洗通过人为制造极端的不确定性，打断了代理人之间的信任构建和契约演化，将系统强制重置为一种低连接度的状态。

2. 强制原子化作为“去熵”校准

清洗通过肉体上的消灭或职权上的剥夺，确保持有中间权力的元胞始终处于一种极度的原子化和受惊吓状态\cite{arendt1973origins}。在这种状态下，个体元胞丧失了进行长程博弈（Long-range collaboration）的能力，唯一的生存策略是向中心表示绝对效忠。

3. “功能性死亡”的循环再生

定期清洗虽然在短期内解决了中心的焦虑，但其长期代价是巨大的：1，每一次清洗都会抹除积累的专业经验与治理信息。2，系统最终只剩下最缺乏能动性、最平庸的执行者，进一步加速了社会向“热寂”状态的坠落。

这种通过清洗维持的有序，是典型的以牺牲系统演化潜能为代价的减熵陷阱。当外部冲击来临时，这种高度原子化的中间阶层将因缺乏自组织韧性而迅速崩解，导致利维坦陷入整体性的系统性崩溃。

科学技术的发展实际上进一步强化了这种格局。在十九世纪末二十世纪初，随着自动机枪的发明，导致一个军人就可以统治管理上千个平民，所以殖民统治达到了高峰\cite{headrick1981tools}。而到了现代，深度自动化（人工智能，互联网，机器人）的发展进一步强化了这种格局。现在借助这些技术，管理效率已经达到登峰造极的地步。而利维坦的管理又十分粗糙，所以利维坦借助技术手段，已经有可能可以做到大幅精简层级，形成一种完全扁平原子化的管理体系了。在这种情况下，官僚阶层面临永久性消失的处境。

除了打击多层结构，利维坦还会打击绝对贫困。如果一个社会有大规模的绝对贫困人口，又有另一群相对不那么贫困的人口，这对利维坦而言也是一种威胁，因为实质上也形成了层级。当地主控制了哪怕仅够一人糊口的口粮，他就能换取流民的绝对效忠。只要存在绝对贫困，中间层级就能通过“施舍生存权”来截断底层对利维坦的直接依附。这种“连饭都吃不起”的流民成为了中间层级的私人武装或极廉价劳动力。这使得中间层级在面对利维坦时，拥有了更强的对抗筹码。而通过将底层人口拉回生存线之上，利维坦实现了切断雇佣动员链：当流民不再需要为了一口饭而自愿奴役于地主时，中间层级的组织资源就会迅速枯竭。底层民众的效忠对象从近端的中间层级转向了远端的利维坦。

从信息论视角看，绝对贫困群体通常是系统的“信息黑洞”。当人处于极端贫困时，其行为高度不可预测且受生物本能驱动，会产生大量的系统噪声（如流民暴动、非法流动），这大大增加利维坦的监控成本 $C_m$。利维坦追求的是一种“均质化的平庸”。利维坦最喜欢的社会形态是——所有人都处于“虽贫穷但能存活”的状态。在这种状态下，没有个体有足够的剩余资源去挑战规则，也没有个体有足够的绝望去掀翻系统。

\section*{制度死锁与转型困境}

虽然单一强人通过垂直整合将系统熵值压缩至局部最小值，但这种秩序并非基于内生协调，而是基于外部压制。这种“静态减熵”在本质上改变了社会的拓扑结构，造成了不可逆的制度损伤。

\subsection*{熵减的代价：耗散结构的停滞}

在非平衡态热力学中，健康的社会系统应表现为一种耗散结构（Dissipative Structure），通过不断引入“涨落”（Fluctuations）来实现更高维度的自组织有序 \cite{prigogine1977self}。

在耗散结构理论中\cite{prigogine1977self}，一个系统要维持低熵（秩序），必须从外部吸取“负熵”并向外排放“熵增”。强人建立的 Monolithic Leviathan 看起来高度有序，但这种秩序是靠强行抑制$L_0$的自发博弈实现的。为了维持这个单一中心的绝对秩序，系统必须消耗巨大的能量来维持监控、镇压和信息过滤。这种能量原本应该用于社会生产，现在却被耗散在维持结构僵化上。系统进入一种“热寂”（Heat Death）预备态。

实质上，通过“制度冻结”，社会的动态无序被转变为静态的僵化。这种“有序”和死亡的有序类似，是以发展彻底停滞为代价的减熵陷阱。而真正的低熵有序（如健康的民主或复杂的生态系统）是靠信息的高效流动维持的；而强人的低熵是靠停止流动维持的。

当这种社会结构发展到极致时，便会表现为大规模的躯体残损（Somatic Mutilation）——即对人体组织的刻意破坏。历史上中国的阉割（Eunuchism）与缠足（Foot-binding），在本质上都是以生命的潜能为代价换取静态秩序的机制。通过使身体的特定部分在功能上“死亡”或丧失流动性，系统人为地强制维持了一种低熵的稳定状态。

值得注意的是，这些习俗的盛行时期与皇权强化的阶段基本重合。这是一个通过物理牺牲来实现“制度冻结”的过程：在此语境下，肉体的残损成为了宏观政治中“减熵陷阱”的隐喻——即局部运动与能动性的停止（功能性死亡），成为了利维坦实现单一化稳定的前提条件\cite{foucault2012discipline}。

\subsection*{自组织能力的丧失与拓扑降维}

强人对中间阶层的“功能阉割”不仅是为了租金归拢，更是为了消除任何潜在的横向协调节点。这种治理逻辑导致社会拓扑结构发生了根本性位移。

\subsubsection*{1. 从“复杂网络”向“星型拓扑”的降维}

原有的多层递归结构虽然低效且高熵，但仍保留了层级间博弈的“残余肌肉”。单一强人通过“去中介化”，将社会结构强制降维为以强人为中心的星型拓扑（Star Topology）。在这种结构下，所有节点（个体）之间被互不信任的信息孤岛所隔离，唯一的连接是与强人的纵向指令流。

\subsubsection*{2. 转型成本}

根据 \cite{acemoglu2006economic} 关于制度转型的论证，成功转型的关键在于社会具备处理冲突的议价能力。然而：

\begin{itemize}\item 

自组织肌肉萎缩： 长期处于“政治强人指令接收”状态的社会，彻底丧失了在下进行自主博弈的记忆。而中间阶层的彻底坍缩，导致去中心化的自组织失去了组织者。

\item 

路径依赖与恐惧： 一旦强人权力松动，由于缺乏中间层级的缓冲和横向契约的保障，原子化的个体基于博弈论的理性预期，会推断系统必然重返“多层掠夺”的混乱状态（$-C$）。\end{itemize}

这种对混乱的极端恐惧（Fear of Disorder）导致了极高的转型门槛 $T$。社会因无法承受短期内自组织算法缺失带来的熵增，只能选择回归利维坦的怀抱。正如 \cite{north2009violence} 所指出，这种在“无序丛林”与“静止利维坦”之间的循环，是缺乏开放准入秩序社会的共同悲剧。

\section*{从对称冲突到层级化的中间态：异质性与局部均衡}

一个关键的问题是：既然单一强人在信息处理和租金提取上最具效率，为何社会不直接从原子化混乱跃迁至单一利维坦，而是必然经历一个痛苦的“多层掠夺”阶段？

本文认为，这是由以下三个逻辑决定的：

1. 权力的刺探成本与物理阈值

在 $L=0$ 的初始状态，没有任何既有的信息基础设施。一个潜在的强人如果想要直接统治 $N$ 个原子化个体，他必须支付 $N \times C_m$ 的监控成本。由于 $N$ 通常巨大，且早期暴力工具的投射半径有限，任何个体都无法在初期实现全局覆盖。因此，权力首先在局部坍缩。局部暴力领袖通过压制周边的生产者形成小型治理单元。这种局部均衡的集合，客观上构成了社会的第一层掠夺网（$L_1$）。

2. 社会异质性与“代理人”的内生化

社会成员在暴力潜能（$C$）和产出效率（$\alpha$）上是高度异质的。当一个顶级强人出现时，次一级的暴力拥有者面临抉择：是与顶级强人进行 $(G, G)$ 这种损耗为 $-C$ 的决战，还是接受“收编”成为中间层级（$L_{k-1}$）？对于顶级强人而言，直接管理底层的成本过高。通过保留中间掠夺者的部分既得利益，让他们担任“承包商”，是一种降低初期管理熵的权宜之计。这种“行政发包制”的逻辑，虽然引入了代理熵，但在物理扩张阶段是唯一可行的路径。

3. 递归结构的稳定性幻觉

在演化早期，多层结构由于实现了权力的局部对齐，会表现出一种伪稳态。每一层的掠夺者都认为自己通过向上缴贡换取了向下掠夺的特许权，从而在局部逃离了霍布斯陷阱。这种“层级分封”的诱惑力使得系统迅速向纵深方向（增加 $k$ 值）而非扁平方向演化。直到这种递归结构延伸到信息熵爆表的临界点，即中间层的寻租损耗抵消了所有生产剩余时，系统才会触发向单一强人的“减熵收缩”。

此外，根据埃莉诺·奥斯特罗姆 (Elinor Ostrom)的观点\cite{ostrom1990governing}，长期共同生活的社群通常建立了复杂的“内部博弈规则”来防止掠夺。所以很容易形成多个权力中心，互相竞争并合作。客观上也容易形成层级化。

\section*{结论：制度进化的死胡同}

本文通过构建多层递归掠夺模型，系统地论证了在规则真空环境下权力结构的演化逻辑。这一演化并非偶然的暴力角力，而是受制于深层的信息学与博弈论规律。

\subsection*{研究结论的系统总结}

根据前述推导，我们得出以下三个核心科学结论：

\begin{enumerate}\item 

结构的收缩必然性： 递归掠夺结构在数学上由于“双重边际化”与“信息熵爆表”而具有本质的不稳定性。权力向单一奇点的收缩，实际上是系统为了对抗层级化租金耗散所引发的自发性减熵行为 \cite{williamson1967hierarchical}。在缺乏法治的条件下，单一强人是系统避免彻底崩解的唯一内生解。

\item 

生存约束的精准锚定： 强人体制的长期稳态（Stationary Bandit）并非仅靠暴力维持，而是源于其对底层生存约束 $\Phi_j(t)$ 的“精准宽容” \cite{olson1993dictatorship}。通过消除中间代理人的随机噪声，强人能够将社会压榨精确地维持在触发布尔冲突的阈值之上。\item 

演化停滞的代价：强制减熵剥夺了社会作为复杂自适应系统的“涨落”与“变异”空间。这种由指令驱动的秩序（Command Order）以牺牲文明的多样性与长期创新动力为代价，使社会陷入一种高压下的静态热寂 \cite{prigogine1977self}。\end{enumerate}

需要注意的是，上述模型都是建立在无规则真空下的。所谓的掠夺规则是强人的“算法优化”，目标是租金最大化。

而真正破局的，就是将掠夺本身定义为非法，在整个社会实现法治。从而彻底推翻上述演化模型。单一强人只是“无规则真空”下的局部最优解（Local Optimum），而法治是通往“全局最优解”（Global Optimum）的唯一跳板。当社会接受了法治，它就从一个“封闭的掠夺系统”转化为了一个“开放的演化系统”。这一跃迁完成了从丛林到文明的升维。

\subsection*{多层结构作为“权力的摩擦力”与法治的孵化器}

在演化制度学视角下，法治社会并非由同质化的原子个体直接构成，而是由大量遵循统一规则的去中心化自组织交织而成。值得关注的是，这些自组织在结构上与“多层掠夺阶段”的掠夺集群具有高度的同构性：它们都拥有内部科层、领导核心以及明确的动员能力。

多层掠夺结构的存在，本质上是在单一强人（Monolithic Leviathan）与底层原子之间制造了“权力的摩擦力”。在多层掠夺阶段，各掠夺集群为了租金最大化，必须在竞争中达成某种关于“势力范围”的界定。这种博弈均衡并非源于对正义的追求，而是谢林（Schelling）所定义的协调博弈（Coordination Games）的产物 \cite{schelling1980strategy}。这种基于实力边界的规则，实质上是一种“停火协议”。

根据威廉姆森（Williamson）的交易成本理论，当多个权力中心并存时，若交互完全依赖随机暴力，系统将面临极高的度量成本与强制执行成本，导致租金被博弈损耗殆尽 \cite{williamson1985economic}。为了降低这种跨层级的博弈内耗，系统会自发涌现出超越个别意志的普遍规则（Universal Rules）。正如诺斯（North）在论述英国光荣革命时所指出的，法治的最初形态往往是精英集团之间为了保护各自租金免受皇权随意剥夺而达成的可信承诺（Credible Commitment） \cite{north2009violence}。

若社会在此微妙的平衡点，因内生或外生变量（如商业伦理的渗透或对长期预期的共识）开始普遍接受法治，便可能完成向现代社会的跃迁。在这种路径中，原先的“掠夺集群”直接转化为法治架构下的“自组织”，其领导者演变为公民社会的精英。这一转型轨迹在英国的男爵议会与日本封建领主向现代精英的转化中得到了充分验证。

法治的核心原则并非纯粹的经验产物，而是基于人类协作博弈的哲学原理或科学规律推导而出的，具有严密性与极高的普适性（Generality）。从系统论的角度看，法治提供了一套低信息熵的元规则（Meta-rules），其在大方向上的确定性确保了不同利益主体之间能够建立长期且稳定的预期。

尽管各地区在执行细节上可能因实际情况（Local conditions）而存在差异，但其底层逻辑是统一的。这种普遍性保证了系统内部的“兼容性”。

相反，若各掠夺方在博弈临界点始终拒绝走向规则的普遍化（Universalization of Rules），误以为维持一种规则真空，或依赖某种基于身份差异、非对称协议的“差异化规则”能维持秩序的稳定，系统将陷入极高的不确定性中。

根据哈耶克（Hayek）的论证，法律的普遍性是自由与秩序的基础，而这种特殊主义规则则属于“命令”范畴 \cite{hayek2020constitution}。从博弈论视角分析，这种规则体系存在两个致命的系统缺陷：

1，解释权的无限寻租：由于缺乏统一的哲学和科学支点，规则的解释权沦为实力的博弈。这导致了谢林（Schelling）意义上的“聚点”不断漂移，系统永远无法达成稳定的纳什均衡 \cite{schelling1980strategy}。

2，黑帮逻辑的内耗陷阱：特殊主义规则建立在人身依附而非契约之上。正如黑帮规矩依赖于头目的个人威望或即时的暴力恐吓，一旦权力中心发生微小的扰动（Fluctuation），规则便会瞬间瓦解。

当规则不再统一，每一场博弈都变成了个案式的权力较量。这导致了交易成本的指数级增长：掠夺者不仅要通过暴力提取租金，还必须消耗巨大的能量来防御其他掠夺者对规则解释权的挑战。这种状态在非平衡态热力学中表现为系统信息熵的剧增。这种以牺牲科学普适性换取的短期稳定，实际上是前述“减熵陷阱”的另一种表现形式。

在这种持续的、高熵的动荡中，由于缺乏普遍规则的锚定，任何微小的力量失衡都会触发正反馈效应。各方会通过打破这种脆弱的“停火协议”，吞噬周边的竞争对手以追求极端的确定性。由于不存在普遍契约（Constitutional Commitment）\cite{north2009violence}，系统最终将不可避免地跨越相变阈值，坍缩为单利维坦格局（Monolithic Leviathan）。

这种坍缩是一种“退化性有序”。它以抹除多层掠夺中所有的局部规则为代价，通过强行降低系统的自由度来压制不确定性。最终，社会走入前述的减熵陷阱：在消灭了博弈摩擦的同时，也彻底扼杀了法治孵化所需的动态演化空间。权力的“去熵”操作导致社会的局部能动性彻底丧失。这种“功能性死亡”导致了个体极度的原子化，使其丧失了自发结社的生物本能。在此语境下，即使单一强人因外部冲击消失，这种丧失了“中间层发育能力”的系统也难以自愈，极易沦为外部发达国家的附庸 \cite{tainter1988collapse}。

\dataavail{ 利维坦演化逻辑的在线展示。
\href{https://lituobang.github.io/Leviathan_Logics.html}{https://lituobang.github.io/Leviathan\_Logics.html} }

\acknow{I acknowledges the Google Gemini in structuring the logic and refining the technical preparation of this work and the simulation online. I would also like to thank the support of peers from UC Berkeley during the preparation of this work.
}

\bibliographystyle{unsrtnat} 
\begin{small}
    \bibliography{references} 
\end{small}

\clearpage
\end{CJK*}
\end{document}